# Genetic Drift and Mutation

Hiroshi Isshiki

College of Science and Technology, Ritsumeikan University, Japan

Abstract: In genetic drift of small populations, it is known that even when the ratio of alleles is 0.5, specific genes are fixed in or disappear from the population. It seems to be the reason why inbreeding is avoided. On the other hand, this phenomenon suggests an interesting possibility. The mutant gene does not increase the number of genes at once in a large population. A gene is partially fixed by increasing the number within a small population because of inbreeding, and the gene increases in a large group by Darwin's natural selection. It would be more reasonable to think in this way. We studied this mathematically based on the concept of genetic drift. This suggested that inbreeding could be useful as a trigger for fixsation of mutation.

1. Introduction

In genetic drift in a small population, it is known that even when the ratio of alleles is 0.5, specific gene is fixed or disappear within the population. This suggests the possibility of disappearance of a favorable gene, for example genes related to fertility, or fixing of an unfavorable gene that inhibit health such as the occurrence of diseases. It seems to be the reason why inbreeding is avoided.

There are interesting studies such as the following [1].

According to Mr. Yoshimaro Yamashina, the founder of Yamashina Institute for Ornithology who thoroughly studied inbreeding of chickens, when repeated inbreeding to birds, abnormal individuals were born until the fifth generation. Except for them, when breeding only excellent individuals, after ten or several generations excellent individuals are born rather than the first one.
There is an effect called the bottleneck or founder effect [2].

In the same principle as the bottleneck effect, only a small part of the population is isolated, and a similar group can be formed when its descendants breed. In this case, because the genotype of a small number of isolated individuals (founder) is inherited, it is called the founder effect.

As a specific example, it can be mentioned that the blood type of Native American is almost O. In this case, it is considered that the O gene frequency became high because many O happens to a very small number of families who crossed the Behring Strait during the Ice Age, and they became the ancestors of all Native Americans. Even if, the O gene frequency of the ancestor who crossed the Behring strait was not dominant, this could happen according to the result of the present paper. The fact that the population was small is the key point to this case.

The phenomenon of the fixsation or disappearance of specific genes in a small group suggests the possibility of giving rise to a chance that a mutation will settle within a population. The fixsation of a mutant gene does not increase the number of genes at once in a population, but a gene partially fixed by increasing the number within a small population due to inbreeding, and it transfers the nature in a large group by Darwin's natural selection. It would be more reasonable to think that fixation progress in this way.

We examine this mathematically based on the concept of genetic floating. We would like to show the possibility that inbreeding will be useful as a trigger for fixation of mutation.

## 2. Genetic drift

With reference to the simulation result of genetic drift described in Ref. [1, 2], simulation method and calculation codes were independently developed.

In order to make simple calculation possible, unnecessary elements were removed as much as possible. We make the following assumption.

(1) We target one allele.

(2) There is no distinction of sex.

(3) A group with population N (even number) creates randomly N / 2 pairs at once.

(4) Each pair creates two children at once. This will be the next generation.

(5) The initial condition is that the ratio of alleles is 0.5 in all individuals, namely (1, 0).

Computation to reproduce the results described in Ref. [3, 4] was performed, and the results shown in Fig. 1 below were obtained. Since these results are random number simulations, they do not completely agree with the results of Ref. [3, 4], but because they are well matched, it can be considered that the result was successfully reproduced. The validity of the calculation of the present paper was secured this way.

In the case of small groups the fact that there are cases where the gene ratio is 1 or 0 is very noticeable. This is pointed out in Ref. [3, 4]. The reason that the gene ratio becomes 1 means that all individual genes are (1, 1), and that the gene ratio becomes 0 means that all individual genes are (0, 0) .

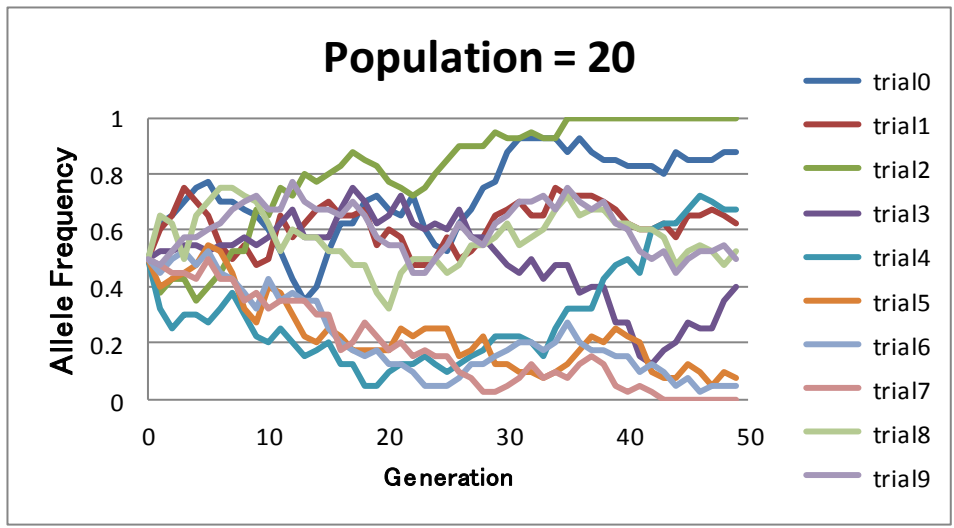

(a) Drift over 50 generations of a small group (Population = 20)

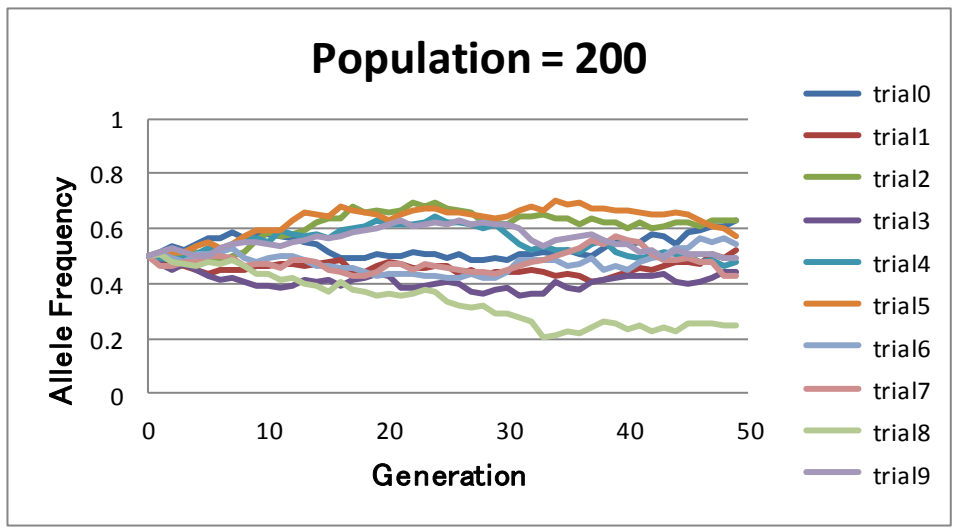

(b) Drift over 50 generations of middle group (Population = 200)

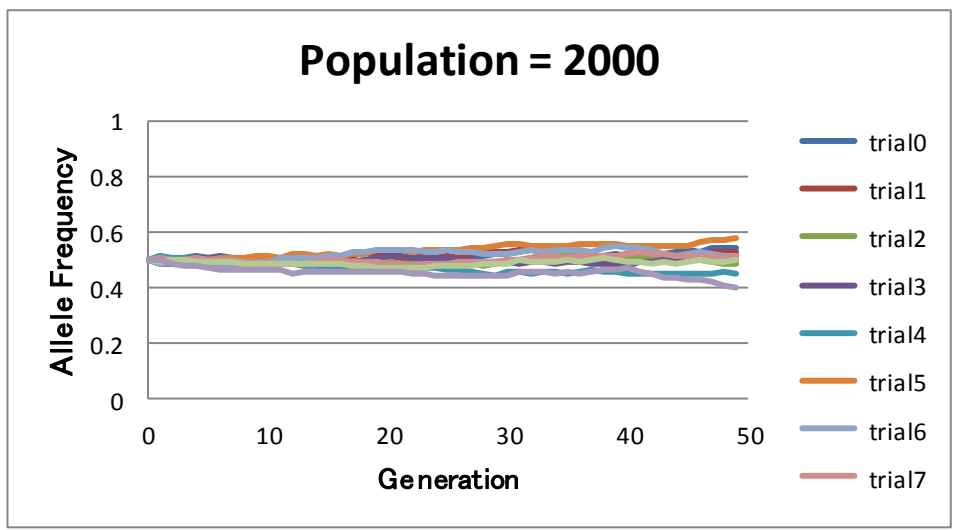

(c) Drift over 50 generations of large group (Population = 2000)

Fig. 1 Genetic drift when the initial ratio of gene is 0.5

## 3. Fixation of mutation

Next, we consider mutations. In the above calculation, the initial value of the gene ratio of each individual was set to 0.5. That is, they are all (1, 0). If this initial value is set to one individual (1, 0) and everything else is (0, 0), it will be considered to correspond to the mutation. This is the only difference from the genetic drift discussed above. For example, if everything becomes (1, 1) after 50 generations, it can be considered that the mutation has settled within the group. In what case is it possible?

Therefore, in the following, we consider the following items:

(1) Population size

(2) Introduction of natural selection factors

## 3.1 Effects of population size

Regarding mutation fixsation, it seems to have the following possibilities.

(1) With respect to mutation fixsation of a small group, it does not necessarily require natural selection.

(2) If fixation of mutation occurs a small group, there is a possibility that they will spread to the whole group and become fixed in the whole group.

(3) If (2) happens, it seems that inbreeding crosses play a major role because there are many inbreeding crosses in a small group.

As you can see, the example of a population consisting of two individuals as shown in Fig. 2 would help your understanding.

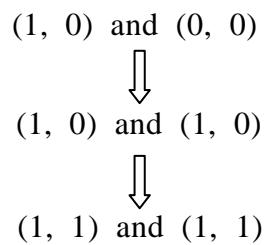

Fig. 2 Generation and fixsation of mutation in two individuals

Simulation was done for a small group to show that mutation is possible without natural selection. Consider a population in which one individual of population N has genes of (1, 0) and all other individuals have genes of (0, 0). For the case of 6 individuals, time series calculation over 50 generations is performed and an example of preliminary time series calculation is shown below.

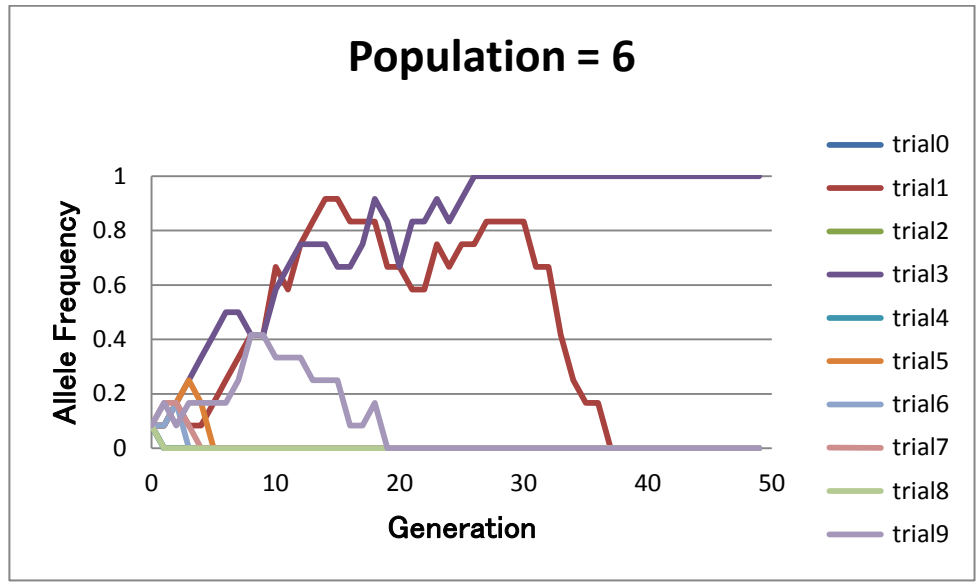

Fig. 2 Calculation example showing the possibility of mutation fixsation

While the population N undergoes 50 generations, the ratio at which genes of all individuals become (1, 1) is the m - ratio (mutation ratio), and if the genes of all individuals are (0, 0),we call the e-ratio (extinction ratio) and the ratio outside of them as d-ratio (drifting ratio). We change the population number N of the group to 2, 4, 6, 10, 20, 40 and tried 30,000 times on each of them. We obtain the numbers of (1,1), (0,0) and others The ratios of the numbers to the total number of trials are defined as m-ratio, e-ratio and d-ratio. Table 1 and Fig. 3 show these results.

Table 1 Mutation fixsation and extinction in small group

| N | m-ratio | e-ratio | d-ratio |
|---|---------|---------|---------|
| 2 | 0.2498 | 0.7502 | 0 |
| 4 | 0.1193 | 0.8688 | 0.0119 |
| 6 | 0.0634 | 0.9014 | 0.0352 |
| 10 | 0.0174 | 0.9129 | 0.0697 |
| 20 | 0.0008 | 0.9237 | 0.0755 |
| 40 | 0 | 0.9267 | 0.0733 |

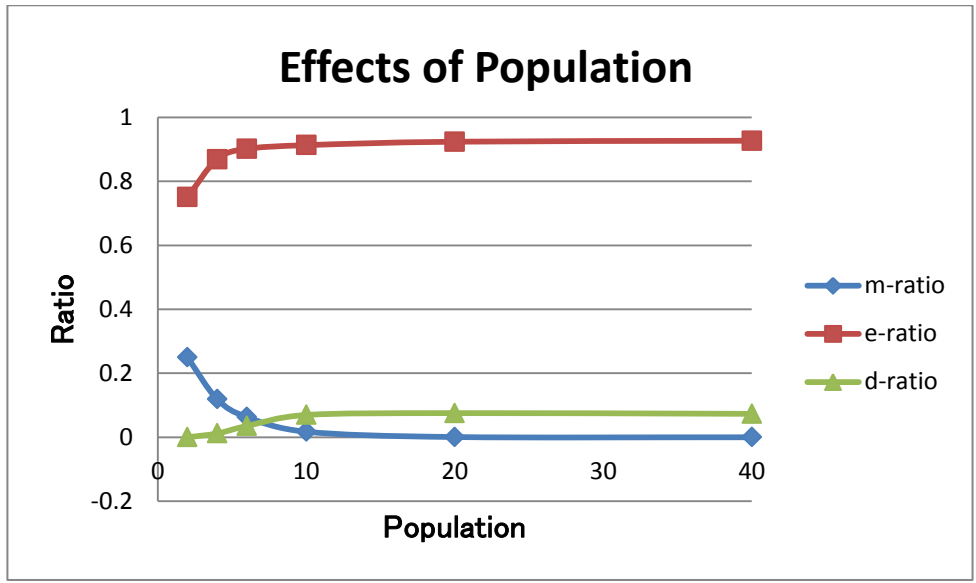

Fig. 3 Mutation fixsation and extinction in small group

When the initial value of one individual gene is (1,1), the same calculation results are shown in Table 2 and Fig. 4. Since the gene of this case is almost double of the case of (1, 0), the nonlinear factor seems to be small.

Table 2 Mutation fixsation and extinction in small group

| N | m-ratio | e-ratio | d-ratio | total |
|---|---|---|---|---|
| 2 | 0.4961 | 0.5038 | 1E-04 | 1 |
| 4 | 0.2407 | 0.7388 | 0.0205 | 1 |
| 6 | 0.1319 | 0.796 | 0.0721 | 1 |
| 10 | 0.0367 | 0.8276 | 0.1357 | 1 |
| 20 | 0.0019 | 0.8464 | 0.1517 | 1 |
| 40 | 0 | 0.8516 | 0.1484 | 1 |

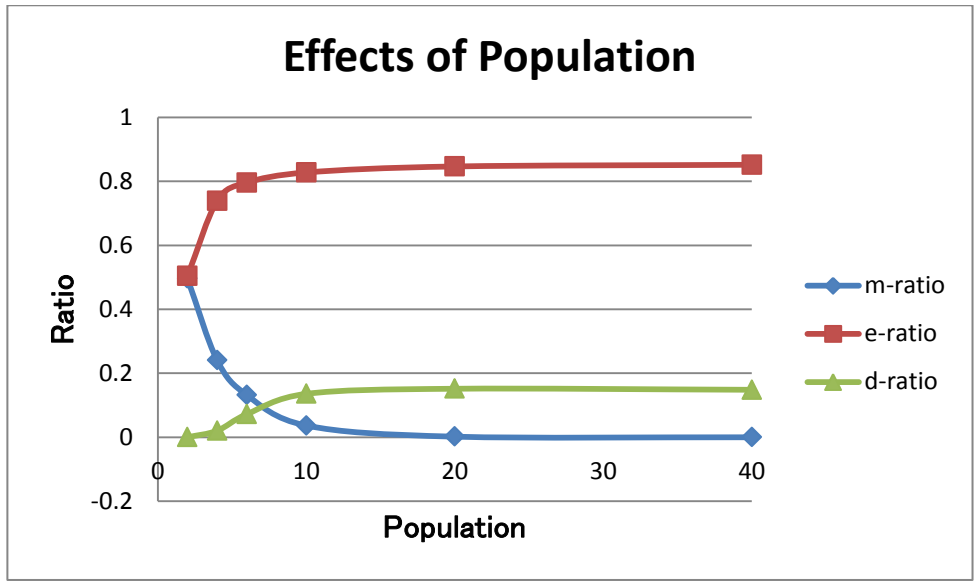

Fig. 4 Mutation fixsation and extinction in small group

### 3.2 Introduction of natural selection factors

An example of introducing a natural selection factor to favor mutation is shown in Fig. 5. In this example, the probability of selecting mutant gene 1 is 0.6 and the probability of selecting gene 0 is 0.4. One individual in the group has genes (1, 0), and the others have genes (0, 0). Compared to Fig. 2, the fixing ratio e-ratio is larger, which shows that fixing is promoted.

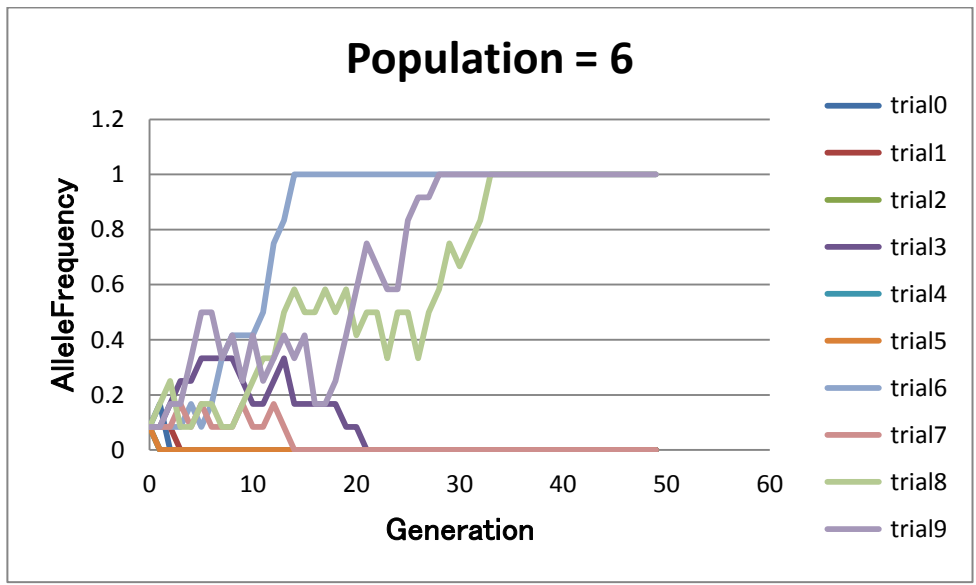

Fig. 5 Calculation example showing the effect of natural selection on mutation fixsation

In Table 3 and Fig. 6, the fixsation ratio m - ratio, the disappearance ratio e - ratio, and the drifting ratio d - ratio are obtained when changing the probability of selecting the mutant gene 1 from 0 to 1.0, with the number of individuals being 20, the number of generations being 50, the number of trials being 30,000. One individual in the population has an initial gene of (1, 0), and genes of all other individuals are (0, 0). There is no natural selection when the probability NS of natural selection is 0.5. There is a positive natural selection when it is larger than 0.5, and a natural selection of minus when it is smaller than 0.5. As the natural selection becomes larger than 0.5, that is, as the plus increases, the fixing ratio m-ratio of the mutation increases and the extinction ratio e-ratio decreases.

Table 2 Mutation fixsation and extinction in small group

| NS | m-ratio | e-ratio | d-ratio |
|---|---|---|---|
| 0 | 0 | 1 | 0 |
| 0.2 | 0 | 1 | 0 |
| 0.4 | 0 | 0.998 | 0.002 |
| 0.425 | 0 | 0.9937 | 0.0063 |
| 0.45 | 0 | 0.981 | 0.019 |
| 0.475 | 0 | 0.959 | 0.041 |
| 0.5 | 0.0006 | 0.9207 | 0.0787 |
| 0.525 | 0.0041 | 0.866 | 0.1299 |
| 0.55 | 0.0177 | 0.7986 | 0.1837 |
| 0.575 | 0.0555 | 0.726 | 0.2185 |
| 0.6 | 0.129 | 0.6599 | 0.2111 |
| 0.8 | 0.754 | 0.2456 | 0.0004 |
| 1 | 1 | 0 | 0 |

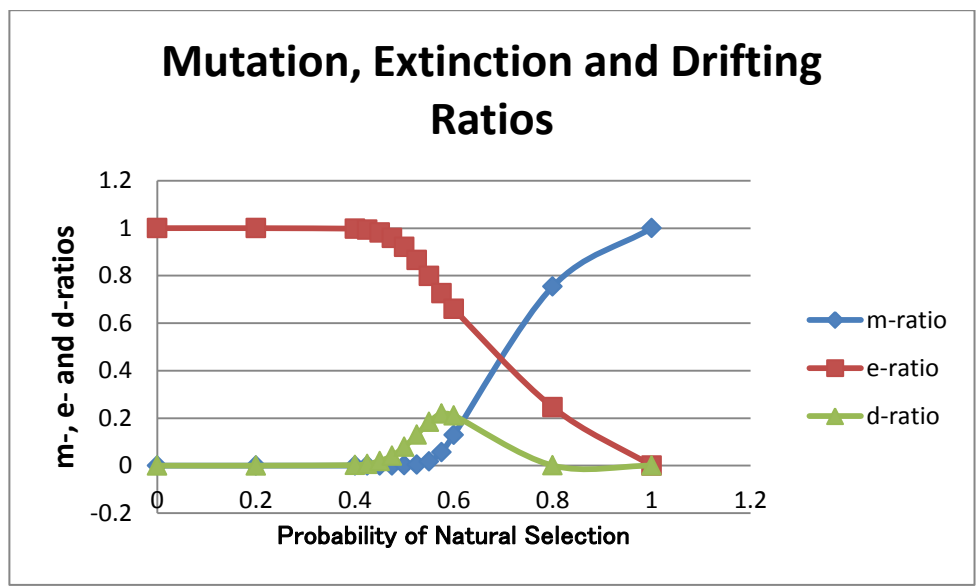

Fig. 6 Mutation fixsation and extinction in small group

## 4. Conclusions

Based on the concept of genetic drift, the mutant genes do not increase the number within a population at once. First, due to inbreeding, the number is increased within the small group and the gene is partially fixed. After that, it is fixed by the natural selection of Darwin among the large group. We simulated and calculated this possibility mathematically. This suggested that inbreeding could be useful as a trigger for fixsation of mutation.

As forms of selection, there may be an artificial compulsion such as breed improvement, an environmental compulsion due to bottleneck as seen in Indian blood type, and one by the Darwin's true natural selection.

When large genetic changes occur, it may lead to a situation leading to the occurrence of new species. Will it happen in one species, or will it be caused by mixing genes beyond species? What is the trigger? It is also interesting problem.

We tried to think about the evolution of the length of the neck of the giraffe in the sense of challenging the established concept. The current giraffe is thought to have evolved from a short neck giraffe to a long neck. However, there seems to be the following mysteries.

(1) If the neck is long, a giraffe has to send blood to the head at a high place against gravity. Therefore, the giraffe has a blood pressure regulating organ called wonder net. However, there are also wonder nets on the fossil of a short neck giraffe too.

(2) Although fossils in the neck of a short giraffe are found, fossils showing a change in neck becoming longer are not found sufficiently [5].

This is regarded as a question of the evolution of the head of Giraffe. If so, will not it be considered that evolution has occurred in the opposite way? In other words, "How was it that Giraffe's neck was originally long, but was mutated and a short neck giraffe was born and temporarily flourished, but it extincted"? As far as the mystery mentioned above is concerned, this would be a reasonable way of thinking.

**Appendix C-language code of genetic drift and natural selection Mutation**

```c
/* ------------------------------------------------------------------ */
/*                                                                    */
/*    File Name: GeneticDrift.c              2018.04.12-2018.04.13    */
/*    File Name: MutationX.c                 2018.04.13-2018.04.13    */
/*    File Name: MutationXTrial.c            2018.04.14-2018.04.15    */
/*    File Name: MutationXTrial11.c          2018.04.15-2018.04.15    */
/*    File Name: MutationXTrial11NS.c        2018.04.15-2018.04.15    */
/*                                                                    */
/*    Mutation                                                        */
/*                                                                    */
/* ------------------------------------------------------------------ */

#include <stdio.h>
#include <stdlib.h>
#include <string.h>
#include <math.h>

#define PI     3.14159265358979323846

void main();
void pushKey();

double drand();                 // ranom real nunmber between -1 and 1
int irand0toN(int);             // ranom integer between 0 to N
int irand0or1(int, int);        // 0 or 1

/* ------------------------------------------------------------------ */

int N;                          // population
int genTtl;                     // total number of generation
int genSel;                     // gene selection ... 0: (1,0), 1: (1,1)
int trialTtl;                   // number of trial
double NS;                      // natural selection

int Crm[100][10001][2];         // chromosome
int Chk[10001];                 // check

double alleleFreq[30001][100];  // allele frequency

int NoOf1;                      // number of mutation
int NoOf0;                      // number of extinction

double AMAT[4001][4001];        // matrix

FILE *fp_inp;                   // pointer of input file
FILE *fp_out;                   // pointer of output file

char InputDataFile[80];         // input file name
char OutputDataFile[80];        // output file name
```

```c
char buf[500];

double tmp[4001];                    //

/* ---------------------------------------------------------------- */

void main()
{
    int i, j, trial;
    int gen, i0, i1;
    int j00, j01;
    int j10, j11;

    srand((unsigned)time(NULL));

    // Input file

    sprintf(InputDataFile, "MutationXTrial11NS_inp.txt");

    if ((fp_inp = fopen(InputDataFile, "r")) == NULL) {
        printf("Failed in Reading Input Data File! ... %s\n", InputDataFile);
        exit(1);
    }

    fscanf(fp_inp, "%s %d", buf, &N);
    fscanf(fp_inp, "%s %d", buf, &genTtl);
    fscanf(fp_inp, "%s %d", buf, &trialTtl);
    fscanf(fp_inp, "%s %d", buf, &genSel);
    fscanf(fp_inp, "%s %lf", buf, &NS);

    fclose(fp_inp);

    printf("N       = %d\n", N);
    printf("genTtl  = %d\n", genTtl);
    printf("trialTtl = %d\n", trialTtl);
    printf("genSel =, %d\n", genSel);
    printf("NS = %12.6f\n", NS);

    pushKey();

    // Output file

    sprintf(OutputDataFile, "MutationXTrial11NS_out.csv");

    if ((fp_out = fopen(OutputDataFile, "w")) == NULL) {
        printf("Failed in Reading Output Data File! ... %s\n", OutputDataFile);
        exit(1);
    }
```

```c
        // input data

        fprintf(fp_out, "N =, %d\n", N);
        fprintf(fp_out, "genTtl =, %d\n", genTtl);
        fprintf(fp_out, "trialTtl =, %d\n", trialTtl);
        fprintf(fp_out, "genSel =, %d\n", genSel);
        fprintf(fp_out, "NS =, %12.6f\n", NS);
        fprintf(fp_out, "\n");

        for (trial = 0; trial < trialTtl; trial++) {

            // initial condition
            if (genSel == 1) {
                Crm[0][0][0] = 1;
                Crm[0][0][1] = 1;
            }
            else {

                Crm[0][0][0] = 1;
                Crm[0][0][1] = 0;
            }
            for (i = 1; i < N; i++) {
                Crm[0][i][0] = 0;
                Crm[0][i][1] = 0;
            }

            alleleFreq[trial] [0] = 0.0;
            for (i = 0; i < N; i++)
                alleleFreq[trial][i] += Crm[0][i][0] + Crm[0][i][1];
            alleleFreq[trial][0] /= 2.0*(N+0.0);;

            printf("gen = %d, allele frequency = %12.6f\n", 0, alleleFreq[trial][0]);

            for (gen = 1; gen < genTtl; gen++) {
                for (i = 0; i < N; i++)
                    Chk[i] = 0;

                for (i = 0; i < N/2; i++) {
                    while (1) {
                        i0 = irand0toN(N);
                        if (Chk[i0] != 1) {
                            Chk[i0] = 1;
                            break;
                        }
                    }
                    while (1) {
                        i1 = irand0toN(N);
```

```
                    if (Chk[i1] != 1) {
                        Chk[i1] = 1;
                        break;
                    }
                }

                j00 = irand0or1(gen, i0);
                j01 = irand0or1(gen, i1);

                j10 = irand0or1(gen, i0);
                j11 = irand0or1(gen, i1);

                Crm[gen][2*i][0] = Crm[gen-1][i0][j00];
                Crm[gen][2*i][1] = Crm[gen-1][i1][j10];

                Crm[gen][2*i+1][0] = Crm[gen-1][i0][j01];
                Crm[gen][2*i+1][1] = Crm[gen-1][i1][j11];
            }

            alleleFreq[trial] [gen] = 0.0;
            for (i = 0; i < N; i++)
                alleleFreq[trial][gen] += Crm[gen][i][0] + Crm[gen][i][1];
            alleleFreq[trial][gen] /= 2.0*(N+0.0);;

            printf("trial = %d, gen = %d, allele frequency = %12.6f\n", trial, gen, alleleFreq[trial][gen]);

        }
    }

    // allele frequency
    fprintf(fp_out, "Allele Frequency\n");

    fprintf(fp_out, "gen, ");
    for (trial = 0; trial < trialTtl; trial++)
        fprintf(fp_out, "trial%d, ", trial);
    fprintf(fp_out, "\n");

    for (gen = 0; gen < genTtl; gen++) {
        fprintf(fp_out, "%d, ", gen);
        for (trial = 0; trial < trialTtl; trial++)
            fprintf(fp_out, "%12.6f, ", alleleFreq[trial][gen]);
        fprintf(fp_out, "\n");
    }
    fprintf(fp_out, "\n");

    // number of mutation and extinction
    NoOf1 = 0;
    NoOf0 = 0;
    for (trial = 0; trial < trialTtl; trial++) {
```

```c
            if (alleleFreq[trial][genTtl-1] == 1.0)
                NoOf1 += 1;
            if (alleleFreq[trial][genTtl-1] == 0.0)
                NoOf0 += 1;
        }

        printf("mutation rate = %12.6f\n", (NoOf1+0.0)/(trialTtl+0.0));
        printf("extinction rate = %12.6f\n", (NoOf0+0.0)/(trialTtl+0.0));
        printf("\n");

        fprintf(fp_out, "mutation rate =, %12.6f\n", (NoOf1+0.0)/(trialTtl+0.0));
        fprintf(fp_out, "extinction rate =, %12.6f\n", (NoOf0+0.0)/(trialTtl+0.0));
        fprintf(fp_out, "\n");

        fclose(fp_out);

        pushKey();
}

/* ---------------------------------------------------------------- */

void pushKey()
{
    printf("\n    Push Return Key! ");
    getchar();
    getchar();
}

/* ---------------------------------------------------------------- */

double drand()
{
    return ((rand()+0.0)/(RAND_MAX+0.0)-0.5)*2.0;
}

/* ---------------------------------------------------------------- */

int irand0toN(int N)
{
    return (int)((drand()+1.0)/2.0*(N+0.0));
}

/* ---------------------------------------------------------------- */

int irand0or1(int gen, int i)
{
    double tmp;

    tmp = (drand()+1.0)/2.0;
```

```
        if (Crm[gen-1][i][0] == 1 && Crm[gen-1][i][1] == 0)
            if (tmp < NS)
                return 0;
            else
                return 1;
        if (Crm[gen-1][i][0] == 0 && Crm[gen-1][i][1] == 1)
            if (tmp < 1.0-NS)
                return 0;
            else
                return 1;
        if (Crm[gen-1][i][0] == 1 && Crm[gen-1][i][1] == 1)
            if (tmp < 0.5)
                return 0;
            else
                return 1;
        if (Crm[gen-1][i][0] == 0 && Crm[gen-1][i][1] == 0)
            if (tmp < 0.5)
                return 0;
            else
                return 1;

        return 0;
}

/* ---------------------------------------------------------------- */
```